\documentclass{aa}
\usepackage[varg]{txfonts}
\usepackage{graphicx}
\usepackage{txfonts}
\usepackage{longtable}
\usepackage{lscape}
\usepackage{booktabs}
\usepackage{version}
\usepackage{multirow}

\begin{document}

\authorrunning{L. Bravi et al.}	
\title{The \textit{Gaia}-ESO Survey: kinematical and dynamical study of four young open clusters} 
	
\author{L. Bravi\inst{1,2}, E. Zari\inst{3}, G.~G. Sacco\inst{2}, S. Randich\inst{2}, R.~D. Jeffries\inst{4}, R.~J. Jackson\inst{4}, E. Franciosini\inst{2}, E. Moraux\inst{5,6}, J. L\'{o}pez-Santiago\inst{7},  E. Pancino\inst{2,8}, L. Spina\inst{9}, N. Wright\inst{4}, F. M. Jim\'enez-Esteban\inst{10}, A. Klutsch\inst{11}, V. Roccatagliata\inst{2}, G. Gilmore\inst{12}, A. Bragaglia\inst{13}, E. Flaccomio\inst{14}, P. Francois\inst{15},  S.~E. Koposov\inst{16}, A. Bayo\inst{17}, G. Carraro\inst{18}, M.~T. Costado\inst{19}, F. Damiani\inst{14}, A. Frasca\inst{11}, A. Hourihane\inst{11}, P. Jofr\'e\inst{20}, C. Lardo\inst{21}, J. Lewis\inst{12}, L. Magrini\inst{2}, L. Morbidelli\inst{2}, L. Prisinzano\inst{14}, S.~G. Sousa\inst{22}, C.~C. Worley\inst{12}, S. Zaggia\inst{23}}
	
\offprints{L. Bravi, \email{bravi@arcetri.astro.it}}
	
\institute{Dipartimento di Fisica e Astronomia, Universit\'{a} degli Studi di Firenze, via G. Sansone 1, 50019 Sesto Fiorentino (Firenze), Italy \and INAF-Osservatorio Astrofisico di Arcetri, largo E. Fermi 5, 50125 Firenze, Italy \and Leiden Observatory, Niels Bohrweg 2, 2333 CA Leiden, The Netherlands \and Astrophysics Group, Keele University, Keele, Staffordshire ST5 5BG, United Kingdom \and Universit\'{e} Grenoble Alpes, IPAG, 38000, Grenoble, France \and CNRS, IPAG, 38000, Grenoble, France \and Department of Signal Theory and Communications, Universidad Carlos III de Madrid 28911, Legan\'{e}s, Madrid, Spain \and ASI Space Science Data Center, via del Politecnico snc, 00133 Roma \and Universidade de S\~{a}o Paulo, IAG, Departamento de Astronomia, Rua do Mat\~{a}o 1226, S\~{a}o Paulo, 05509-900 SP, Brasil \and Departmento de Astrof\'{i}sica, Centro de Astrobiolog\'{\i}a (INTA-CSIC), ESAC Campus, Camino Bajo del Castillo s/n, E-28692 Villanueva de la Ca\~nada, Madrid, Spain \and INAF - Osservatorio Astrofisico di Catania, via S. Sofia 78, 95123 Catania, Italy \and Institute of Astronomy, University of Cambridge, Madingley Road, Cambridge CB3 0HA, United Kingdom \and INAF - Osservatorio Astronomico di Bologna, via Gobetti 93/3, 40129, Bologna, Italy \and INAF - Osservatorio Astronomico di Palermo G. S. Vaiana, Piazza del Parlamento 1, 90134 Palermo, Italy \and GEPI, Observatoire de Paris, CNRS, Universit\'e Paris Diderot, 5 Place Jules Janssen, 92190 Meudon, France \and McWilliams Center for Cosmology, Department of Physics, Carnegie Mellon University, 5000 Forbes Avenue, Pittsburgh, PA 15213, USA \and Instituto de F\'{i}sica y Astronomi\'{i}a, Universidad de Valparai\'{i}so, Chile \and Dipartimento di Fisica e Astronomia, Universit\`{a} di Padova, Vicolo dell'Osservatorio 3, 35122 Padova, Italy, \and Departamento de Did\'{a}ctica, Universidad de C\'{a}diz, 11519 Puerto Real, C\'{a}diz, Spain \and N\'ucleo de Astronom\'{i}a, Facultad de Ingenier\'{i}a, Universidad Diego Portales, Av. Ej\'ercito 441, Santiago, Chile \and Laboratoire d'astrophysique, Ecole Polytechnique F\'ed\'erale de Lausanne (EPFL), Observatoire de Sauverny, CH-1290 Versoix, Switzerland \and Instituto de Astrof\'isica e Ci\^encias do Espa\c{c}o, Universidade do Porto, CAUP, Rua das Estrelas, 4150-762 Porto, Portugal \and INAF - Padova Observatory, Vicolo dell'Osservatorio 5, 35122 Padova, Italy}
	
\date{Received --- / Accepted ---}
	
\abstract {The origin and dynamical evolution of star clusters is an important topic in stellar astrophysics. Several models have been proposed to understand the formation of bound and unbound clusters and their evolution, and these can be tested by examining the kinematical and dynamical properties of clusters over a wide range of ages and masses.} {We use the \textit{Gaia}-ESO Survey products to study four open clusters (IC 2602, IC 2391, IC 4665, and NGC 2547) that lie in the age range between 20 and 50 Myr.} {We employ the gravity index $\gamma$ and the equivalent width of the lithium line at 6708 $\AA$, together with effective temperature $\rm{T_{eff}}$, and the metallicity of the stars in order to discard observed contaminant stars. Then, we derive the cluster radial velocity dispersions $\sigma_c$, the total cluster mass $\rm{M}_{tot}$, and the half mass radius $r_{hm}$. Using the \textit{Gaia}-DR1 TGAS catalogue, we independently derive the intrinsic velocity dispersion of the clusters from the astrometric parameters of cluster members.} {The intrinsic radial velocity dispersions derived by the spectroscopic data are larger than those derived from the TGAS data, possibly due to the different masses of the considered stars. Using $\rm{M}_{tot}$ and $r_{hm}$ we derive the virial velocity dispersion $\sigma_{vir}$ and we find that three out of four clusters are supervirial. This result is in agreement with the hypothesis that these clusters are dispersing, as predicted by the "residual gas expulsion" scenario. However, recent simulations show that the virial ratio of young star clusters may be overestimated if it is determined using the global velocity dispersion, since the clusters are not fully relaxed.}{}
	
\keywords{stars: pre-main sequence -- stars: kinematics and dynamics -- open cluster and associations: individual: IC 2602, IC 2391, IC 4665, NGC 2547 -- stars: formation -- techniques: spectroscopic -- techniques: radial velocities} 

\maketitle

\section{Introduction}

The majority of stars form in clusters and associations inside giant molecular clouds. However, most clusters dissipate within 10 -- 100 Myr, leaving more than 90\% of the stellar population dispersed in the Galactic field \cite[e.g.,][]{Lada_2003,Piskunov_2006}. The scientific debate on the origin of bound and unbound clusters, along with the processes leading to their dissolution, is still open. Several authors suggest that all stars form in dense clusters (density $\rm{ \gtrsim 10^{3} - 10^{4}~stars~pc^{-3}}$), which rapidly dissipate after feedback from massive stars (i.e., supernova explosions, stellar winds, and radiation pressure) sweeps out the gas that was keeping the cluster bound \citep[e.g.,][]{Tutukov_1978,Lada_1984,Goodwin_1997,Kroupa_2001, Goodwin_2006, Baumgardt_2007, Bastian_2011}. These models predict that clusters -- after gas dispersion -- should be found in a supervirial state. Recent observations and simulations question this scenario suggesting that clusters have origin in a hierarchically structured environment covering a large range of densities and that the stellar feedback and gas expulsion are irrelevant for the cluster dispersion, which is, instead, driven by two-body interactions \citep[e.g.,][]{Bressert_2010,Kruijssen_2012, Parker_2013,Wright_2016,Parker_2016}. 

In order to achieve a full understanding of the origin and the fate of star clusters, it is fundamental to study the kinematic properties of their stellar components at different stages of evolution. However, until a few years ago  this kind of studies had been carried out only for a few clusters \citep[e.g.,][]{Cottaar_2012_1, Tobin_2015}, due to the lack of precise and homogeneous measurements of radial velocities and other stellar parameters for large stellar samples.
\begin{table*}[!h]
	\centering
	\small{
		\caption{Cluster properties.}
		\label{tab:clusters}  
		\begin{tabular}{cccccc}
			\midrule
			\midrule		
			Cluster & RA & DEC & Distance & Age & E(B-V) \\ 
			& (J2000) & (J2000)& (pc)&(Myr) & \\ 
			\midrule
			IC 2602 & 10h 40m 48s & -64d 24m 00s & $148.0^{+7.3}_{-6.1}$  & $43.7^{+4.3}_{-3.9}$   & $0.068 \pm 0.025$\\[1ex]
			IC 2391 & 08h 40m 32s & -53d 02m 00s & $146.0^{+7.1}_{-6.1}$ & $51.3^{+5.0}_{-4.5}$  & $0.088 \pm 0.027$\\[1ex]
			IC 4665 & 17h 46m 18s & +05d 43m 00s & $366.0^{+46.8}_{-37.9}$  & $23.2^{+3.5}_{-3.1}$   & $0.226 \pm 0.080$\\[1ex]
			NGC 2547 & 08h 10m 00s & -49d 12m 00s & $364.0^{+46.8}_{-37.9}$ & $37.7^{+5.7}_{-4.8}$ & $0.080 \pm 0.024$\\
			\midrule
	\end{tabular}}
\end{table*}
The observational scenario has radically changed very recently, thanks to large high-resolution spectroscopic surveys like APOGEE \citep{Majewski_2017} and the \textit{Gaia}-ESO Survey \citep[GES,][]{Gilmore_2012, Randich_2013}. The latter is a large public survey of all the Milky Way components performed with the multi-object optical spectrograph FLAMES at the Very Large Telescope (VLT). One of the main goals of the survey is the observations of several clusters in the 1 -- 100 Myr age range to derive radial velocities (RVs) and stellar parameters that can be used to investigate their dynamical evolution. \\
\indent
Several interesting results have already been obtained from the first clusters that have been observed ($\rho$ Oph, Chamaeleon I, Gamma Velorum); namely, the discovery of multiple stellar kinematical populations \citep{Jeffries_2014J,Sacco_2015, Mapelli_2015}, and a significant discrepancy between the kinematic properties of pre-stellar cores and pre-main sequence stars formed in the same environment \citep{Foster_2015,Rigliaco_2016, Sacco_2017}. \\
\indent
So far, all these studies have focused on clusters younger than 10 -- 20 Myr. Nevertheless the complete understanding of the cluster dispersion process requires the study of slightly older (age $\sim$ 20 -- 50 Myr) systems. Clusters in this age range have already lost their residual gas and have nearly completed the process of "violent relaxation" predicted by models based on stellar feedback \citep[e.g.,][]{Goodwin_2006,Proszkow_2009}, but have not yet been affected by tidal effects due to external gravitational field that occur on longer timescales \cite[$\sim$ 100 -- 300 Myr; e.g.,][and reference therin]{Portegies-Zwart_1998,Baumgardt_2003,Lamers_2005,Portegies-Zwart_2010}.\\
\indent
In this paper we will investigate this particular age interval using the GES data to analyze the kinematical and dynamical properties of IC 2602, IC 2391, IC 4665, and NGC 2547. 

The paper is organized as follows: in Sect. 2 we describe the observations and the GES parameters used in this paper; in Sect. 3, we illustrate the properties of these clusters and the target selection; in Sect. 4 we explain how we derived the kinematical properties of these clusters; in Sect. 5 we discuss our results; and in Sect. 6 we draw our conclusions.

\section{\textit{Gaia}-ESO observations and data}

GES is obtaining medium and high resolution optical spectra of $\sim$ 10$^5$ stars selected in the Galactic field and in star clusters in order to provide a homogeneous overview of the distributions of kinematics and chemical element abundances in the Galaxy. Specifically, GES is collecting a big dataset of radial velocities (RVs), stellar parameters (i.e., effective temperature, surface gravity, metallicity), and elemental abundances for large numbers of representative stars in clusters, covering a wide range of ages and stellar masses.

\textit{Gaia}-ESO observations are performed with the FLAMES instrument \citep{Pasquini_2002}, using both GIRAFFE and UVES spectrographs, that permit the simultaneous allocation of 132 and 8 fibres, respectively. In the observations of young nearby open clusters, GIRAFFE is used for late-type stars with a V magnitude between 11 and 19 with the HR15N setup, that obtains medium resolution spectra (R $\sim$ 17000) in the wavelength range 6470 $\AA$ < $\lambda$ < 6790 $\AA$. UVES acquires higher resolution spectra (R $\sim$ 47000) of brighter stars (9 < V < 15) with a spectral range of 2000 $\AA$ and with two central wavelengths, 5200 $\AA$ (UVES 520) and 5800 $\AA$ (UVES 580). Both GIRAFFE/HR15N and UVES/580 setups contain the lithium line at 6708 $\AA$, that is useful for identifying young stars.

Pipeline reduction of GIRAFFE spectra and RV determination are centralized at the Cambridge Astronomy Survey Unit (CASU), while UVES reduction and RV analysis are performed at INAF--Osservatorio Astrofisico di Arcetri. The data reduction is described in \cite{Jeffries_2014J} and \cite{Sacco_2014} for GIRAFFE and UVES data, respectively. The reduced spectra are then analyzed using common methodologies to produce a uniform set of stellar parameters, which along with RVs, is periodically released to all the members of the GES consortium via a science archive\footnote{The GES science archive is run by the Royal Observatory of Edinburgh. More informations on the archive are available at the website ges.roe.ac.uk}. 

Spectrum analysis is distributed among several working groups (WGs) and several nodes. WG12 analyzes the pre-main sequence (PMS) stars and different nodes contribute to provide estimates of the stellar parameters and chemical abundances: specifically, two nodes analyze GIRAFFE targets (INAF--Osservatorio Astrofisico di Catania (OACT) and INAF--Osservatorio Astronomico di Palermo (OAPA)), and four nodes focus on the UVES targets (OACT, Centro de Astrofisica de Universidade do Porto (CAUP), Universidad Complutense de Madrid (UCM) and INAF--Osservatorio Astrofisico di Arcetri). The products delivered by the nodes are combined to produce the recommended set of measurements provided by WG12 \citep{Lanzafame_2015}, which in turn is homogenized with those from WG10 and WG11 (GIRAFFE and UVES analysis of FGK stars, respectively) in order to produce the final recommended values reported in the tables \cite[][Hourihane et al., in preparation]{Pancino_2017}. 

During this work, we make use of the RV, the effective temperature of the star ($\rm{T_{eff}}$), the surface gravity (log \textit{g}), the gravity index ($\gamma$), the equivalent width of the lithium line at 6708 $\AA$ (EW(Li)), and the metallicity ([Fe/H]). The $\gamma$ index is an efficient gravity indicator for the GIRAFFE targets when it is combined with the effective temperature of the stars. It is an empirical index and it is sensitive to stellar gravity over a wide range of spectral types, allowing a clear separation between the low gravity giants and the higher gravity main-sequence (MS) and PMS stars for spectral types later than G \cite[see][for details]{Damiani2014}. Given that the measurement of $\gamma$ are available for a larger number of GIRAFFE spectra than the log \textit{g} parameter, we use this as gravity indicator with the exception of the stars observed only with UVES, when $\gamma$ is not derived. For the latter we use instead the value of log \textit{g}, which is available for most of the observed sources.\\
The RVs for the GIRAFFE targets were obtained as explained in \cite{Jackson_2015}, while RVs from UVES are described by \cite{Sacco_2014}. The uncertainties on the RV measurements for GIRAFFE have been calculated empirically using the formula described in \cite{Jackson_2015}, in which they compared repeated measurements of the RV for the same star to determine the underlying distribution of measurement uncertainties as a function of signal-to-noise ratio (SNR), $T_{\rm eff}$ and rotational broadening ($v\,\rm{sin}\, \textit{i}$). In this paper, we use the data from the fourth internal data release (GESviDR4). The values of $v\,\rm{sin}\, \textit{i}$ in recommended table are not available. Therefore, we use the measurements of $v\,\rm{sin}\, \textit{i}$ given by WG12. In a number of cases, recommended values of EW(Li) and $\rm{T_{eff}}$ from the final homogenization process were not provided. In some of these cases we use the EW(Li) and $\rm{T_{eff}}$ derived by the nodes of WG12. This choice is justified by the fact that the values measured by different nodes are in agreement, within the errors, with those recommended by WG12, when there is. 

\section{Sample clusters}

\subsection{Cluster properties}

The four clusters have similar ages (from $\sim 20$ Myr to $\sim 50$ Myr) and different distances. IC 2602 and IC 2391 are among the closest clusters to the Sun (distance $\sim 150$ pc), while the other two clusters are more distant ($\sim 365$ pc). Given the uniform magnitude limit for the observations of clusters in GES, we reach stars with different mass limits in the different clusters (see next Section). The cluster properties are summarized in Table \ref{tab:clusters}, where distances, ages and reddening values are given by \cite{Randich_2017}. Each cluster has been subject to a variety of studies carried out to identify the stellar population based on combinations of X-ray data \cite[e.g.,][]{Prosser_1993,Randich_1995,Patten_1996,Martin_1997,Jeffries_1998}, optical photometry \cite[e.g.,][]{Prosser_1996,Jeffries_2004}, and optical spectroscopy \cite[e.g.,][]{Randich1997,Stauffer_1997,Barrado_1999,Jeffries_2000,Jeffries_2005,Platais_2007,Manzi_2008,Jeffries_2009}; many high and low-mass cluster members were identified using the position in the HR diagram, presence of the lithium absorption line at 6708 $\AA$, and RVs. These studies show that the number of previously known spectroscopically confirmed members in the four clusters range from 40 in IC 4665 to 75 in NGC 2547. In the case of NGC 2547 \cite{Sacco_2015} found a secondary population that is kinematically distinct from the main cluster population.

\subsection{Target selection}
\label{target_selection}
\begin{table}[!b]
	\caption{Number of target observed in the four clusters.}
	\label{tab:selection_cluster}  
	{\centering
		\tiny{
			\begin{tabular}{lccccccc}
				\midrule
				\midrule		
				Name& Setup &N.& N. & N. & N. & N. &  N. \\ 
				& &stars & $\rm{T_{eff}}$ & $\gamma$ &  Log $g$ & EW(Li) &  RV \\
				\midrule
				IC2602 & HR15N  &1528 & 1483 & 1481 & 729 & 1374 &1528\\
				& U 580  & 42 & 42 & - & 42 & 25 & 41\\
				& U 520 & 7 & 7 & - & 7 & - & 6 \\
				& Tot. & 1577 & 1532 & 1481 & 778 & 1399 &  1575\\
				\midrule
				IC2391 & HR15N & 403 & 385 & 378 & 180 & 386 &  402\\
				& U 580 & 20 & 20 & - & 20 & 13 &  20\\
				& U 520 & 8 & 8 & - & 8 & - & 7 \\
				& Tot.&  431 & 413 & 378 & 208 & 399 &  429\\
				\midrule
				IC4665 & HR15N & 545 & 527 & 520 & 258 & 503 &  546\\
				& U 580&  22 & 21 & - & 21 & 19 & 21\\
				& U 520 & - & - & - & - & - & - \\
				& Tot. & 567 & 548 & 520 & 279 & 522 &  567\\
				\midrule
				NGC2547 & HR15N & 450 & 399 & 383 & 149 & 385 &  450\\
				& U 580 & 5 & 5 & - & 5 & 3 &  5\\
				& U 520 & 19 & 18 & - & 18 & - &  13\\
				& Tot. & 474 & 422 & 383 & 172 & 388 & 468\\
				\midrule
				\midrule
				\multicolumn{8}{l}{
					\begin{minipage}{0.48\textwidth}
						\footnotesize{\textbf{Notes.} The table shows the number of targets for which values of different stellar parameter recommended are available (in the case of EW(Li) or $\rm{T_{eff}}$ also from the nodes).}
				\end{minipage}}					
	\end{tabular}}}
\end{table}

One of the main goals of GES is the study of cluster kinematics and dynamics based on large, unbiased samples of members. Known members from the literature do not provide suitable samples, because they are often biased by the selection method. For this reason GES adopts an inclusive selection strategy: all candidate members observed with GIRAFFE have been selected in an unbiased way, down to 19th magnitude (V band), and covering a relatively large area on the sky, from a strip around the cluster sequence. This is defined as the sequence drawn by the known members reported in the literature in the different color-magnitude diagrams (CMDs). When the optical photometry catalogues are either inhomogeneous or incomplete, the selection is based mainly on the photometry of the Two Micron All Sky Survey \cite[2MASS,][]{Skrutskie_2006}. We note that this strategy implies that our final candidate samples include a very large number of fore-/back-ground stars. Inside the magnitude range and spatial coverage observed by GES, some  of the samples are relatively complete; however, in nearby and extended clusters, like those analyzed in the present paper, the level of completeness is lower. Whilst one needs to correct for this incompleteness, our strategy of target selection ensures that the final samples are unbiased (in particular with respect to the kinematics) and representative of the entire cluster population. \\
UVES targets are mainly observed to derive the cluster chemical pattern \citep{Spina_2014b,Spina_2014a,Spina_2017} and are therefore selected with a different strategy: namely, when information is available, the UVES fibres are assigned to brighter stars that are already known or likely members.

In the case of IC 2602 ESO archival data have also been retrieved and analyzed. In order to be consistent with the \textit{Gaia}-ESO selection method, we considered only the archive data for stars that are in the strip of the CMD used for the GES selection.

Table \ref{tab:selection_cluster} summarizes the number of targets observed in each cluster. We list the number of stars observed with the different GIRAFFE and UVES setups, as well as the number of targets for which stellar parameters were derived.
\begin{figure*}[!ht]
	\centering
	\includegraphics[width=0.99\linewidth]{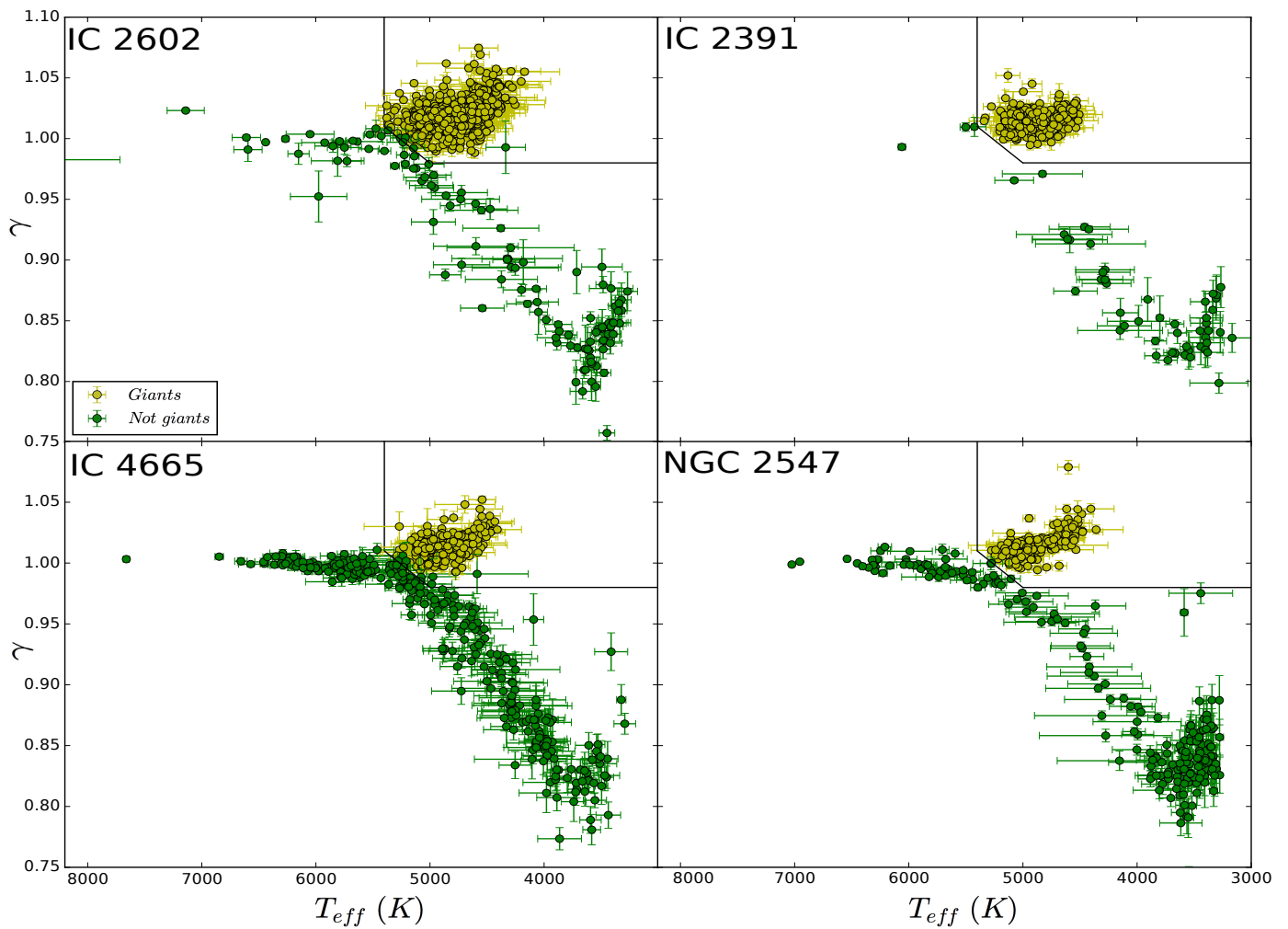}
	\caption{Gravity index $\gamma$ as a function of the stellar effective temperature, $\rm{T_{eff}}$, for the stars observed with GIRAFFE. The yellow filled dots are the stars identified as giants, while the remaining stars are denoted with green filled dots. The black line indicates the threshold used to separate the giants and the non-giants.}
	\label{fig:gamma}
\end{figure*}

\subsection{Completeness}
\label{sec:completeness}

As discussed in the previous Section, the initial targets were selected in order to be complete within the magnitude range of GES and within the area that contains all the stars selected as initial candidate targets, that is defined by the radius $\rm{R_{GES}}$. Therefore, the level of completeness within the observed magnitude range is calculated by dividing the number of the observed stars by the number of stars selected as initial candidate targets in GES and located within these circular regions, which we assume to contain all the cluster. We obtain as level of completeness $\sim 25\%$ for IC 2602 and IC 2391, while for IC 4665 and NGC 2547 we derived a level of completeness of $\sim 65\%$ and $\sim 75\%$, respectively. We note that the level of completeness of IC 2602 and IC 2391 is much because only part of the area of the sky including known cluster members from the literature has been observed.
\begin{figure}[!h]
	\centering
	\includegraphics[width=0.955\linewidth]{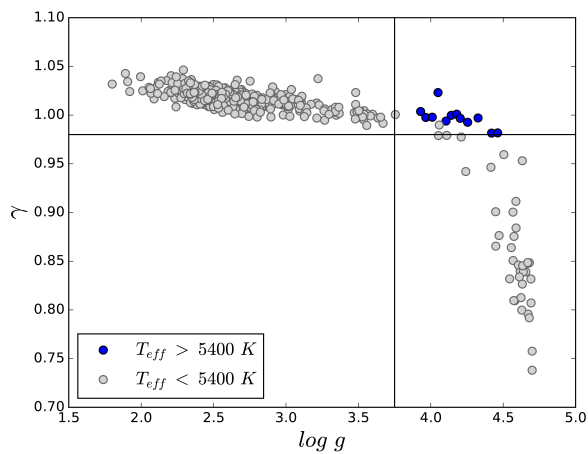}
	\caption{Gravity index, $\gamma$, as a function of the surface gravity, log \textit{g}, for IC 2602. The blue dots are the targets with an effective temperature greater than 5400 K while the grey ones are those with a $\rm{T_{eff}}$ < 5400 K. The solid lines delimit the regions in which a target is considered a giant star based on its $\gamma$ index (above the horizontal black line) and on its log \textit{g} value (to the left of the vertical black line).}
	\label{fig:test_gamma_logg}
\end{figure}

\section{Membership analysis}
\begin{figure*}[!h]
	\centering
	\includegraphics[width=0.98\linewidth]{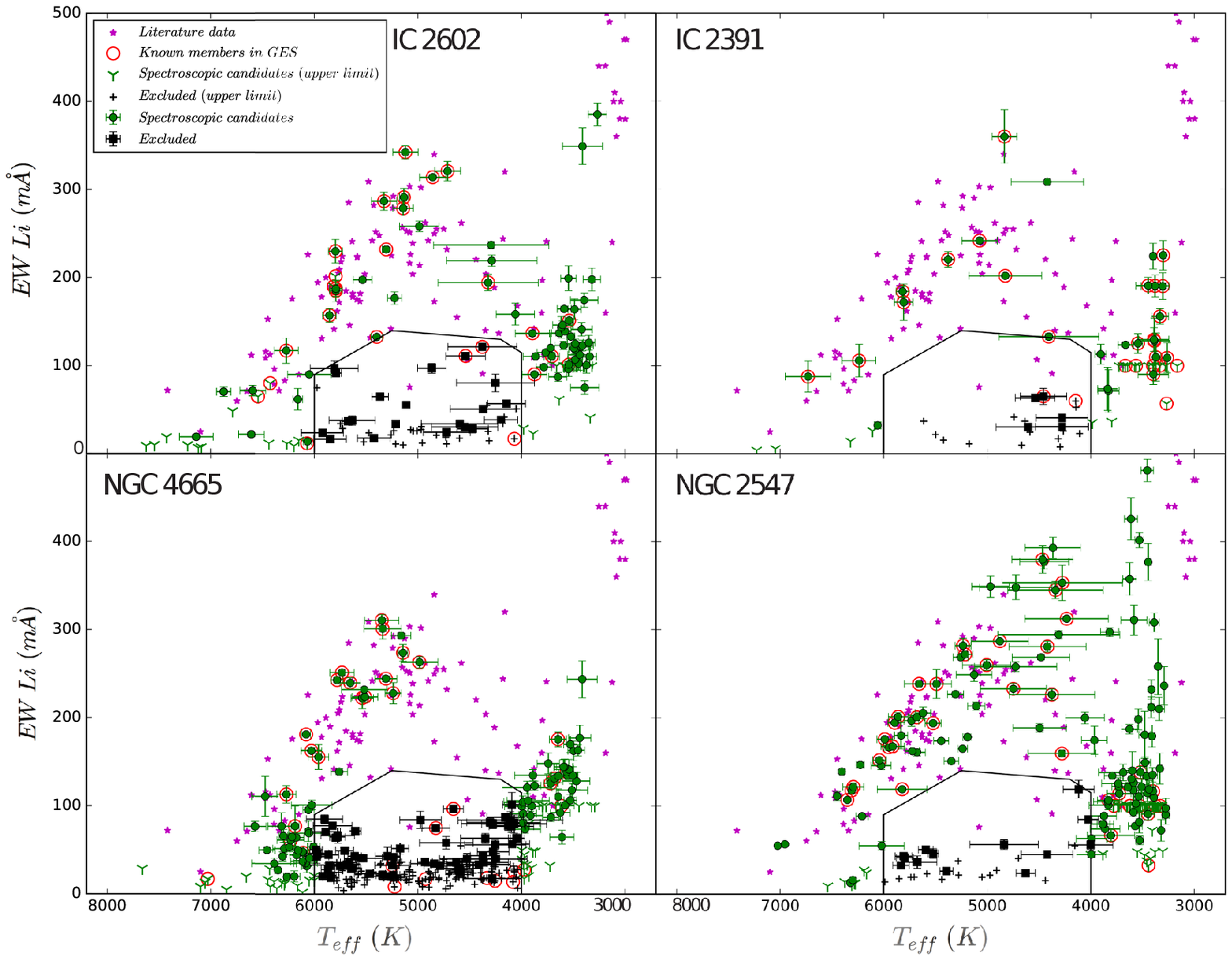}
	\caption{Equivalent width of the lithium line as function of $\rm{T_{eff}}$. The purple star symbols are the known members of the four clusters used to trace the lower envelope of the bulk of cluster spectroscopic candidates (solid line), as given in the literature. The open red circles show the members of each cluster, which have been previously reported in the literature and observed again by GES. The green symbols mark the stars selected as spectroscopic candidates. The stars excluded with the lithium criterion are displayed in black. We also highlight the sources with an estimate of EW(Li) (dots and squares) or only an upper limit (arrows and crosses).}
	\label{fig:litio}
\end{figure*}

Starting from the initial sample of observed cluster targets, thanks to the spectroscopic parameters derived by the GES consortium, we are able to exclude stars that do not belong to the clusters. Then, using the RV of the spectroscopically selected candidates, we can determine a probability that each remaining star is a cluster member and use the Hertzsprung-Russell (HR) diagram to estimate a mass for that star.

\subsection{Spectroscopic candidates}
\label{sec:member}

To exclude stars that do not belong to the clusters we used three independent spectroscopic criteria based on the gravity index $\gamma$ (or log \textit{g} for UVES spectra), the EW(Li), and the metallicity [Fe/H]. All stars where any of the first two parameters or the effective temperature have not been measured have been excluded. We retained stars without the metallicity measurement since very few stars were discarded on the basis of this parameter. More specifically, our method can be divided into three steps.

\begin{itemize}
	\vspace{0.4cm}
	\item[$\bullet$] The main source of contamination in a sample of candidate members of a nearby young cluster are the background giants. These objects have a lower gravity than cluster members and can be identified using the surface gravity index $\gamma$ \cite[]{Damiani2014}. Figure \ref{fig:gamma} shows $\gamma$ as a function of the effective temperature. We consider as giants all the stars within the region defined by the black line, which has approximately $\rm{T_{eff}}$ lower than 5400 K and $\gamma$ higher than 0.98, within one sigma error bar. For UVES targets the gravity index is not defined and we use the surface gravity: we consider as giants stars with log \textit{g} lower than 3.75. In order to check the consistency between using $\gamma$ or log \textit{g}, we plot in Fig. \ref{fig:test_gamma_logg} the comparison between these two parameters for IC 2602: it is clear that the selection of the giants is basically the same whether we consider targets with $\gamma$ > 0.98 and $\rm{T_{eff}}$ < 5400 K or targets with log \textit{g} < 3.75. Indeed, all the targets identified as giants with $\gamma$ and $\rm{T_{eff}}$ are distinctly below the value of log \textit{g} = 3.75. 
	
	\vspace{0.4cm}
	\item[$\bullet$] We use the EW(Li) to exclude dwarf non-members from the sample of stars remaining from the first selection step. Depending on stellar mass, lithium starts to being depleted during the PMS phase \cite[e.g.,][]{Soderblom2010}, therefore it can be used as indicator of youth in specific temperature ranges. Specifically, between 20 and 50 Myr, the EW(Li) can be used to select candidate members between 4000 K and 6000 K because stars with $\rm{T_{eff}} <$ 4000 K have already burned all their lithium and above 6000 K Li is preserved even in much older stars. Also the Li~{\sc i}~6708\AA\ line, that is the main diagnostic, becomes very weak and difficult to measure. In Fig. \ref{fig:litio}, we show the EW(Li) as a function of $\rm{T_{eff}}$ for the four clusters. We classify as secure non-members all the stars below the threshold reported with a continuous black line between 4000 K and 6000 K. All the other stars are selected as candidate members. The threshold has been defined using previous observations of these four clusters available in the literature \citep{Martin_1997,Randich1997,Randich2001,Jeffries_2003,Jeffries_2005,Jeffries_2009}.
	
	\vspace{0.4cm}
	\item[$\bullet$] A final selection step is to exclude targets with a measured [Fe/H] < - 0.5 dex that would be incompatible with the nearly-solar metallicity of these clusters \citep{Spina_2017}.
\end{itemize}

\noindent
To summarize, we retain from the criteria 101, 53, 121, and 187 stars for IC 2602, IC 2391, IC 4665, and NGC 2547, respectively. We define these stars \textquotedblleft spectroscopic candidates''. Given the different target selection strategy used for UVES targets, we also consider as spectroscopic candidate UVES stars without $\rm{T_{eff}}$, $\gamma$, and/or EW(Li) that are known members from the literature. 
  
\subsection{Kinematic analysis}
\label{sec:RV}
\begin{figure*}[!h]
	\centering
	\includegraphics[width=0.95\linewidth]{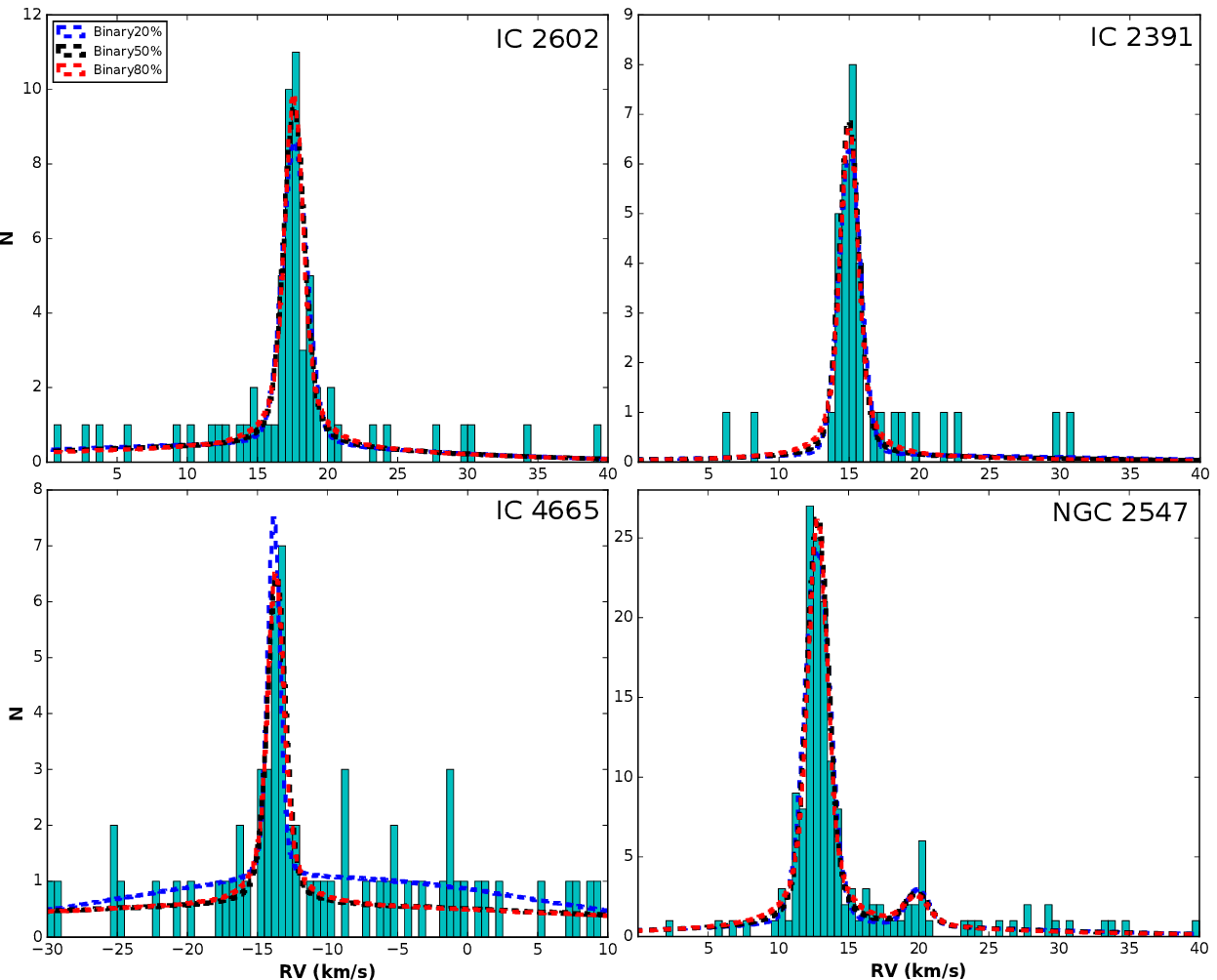}
	\caption{Radial velocity distribution. The blue, black and red dashed lines represent the fits performed by setting the binary fraction to 0.2, 0.5, and 0.8y, respectively.}
	\label{fig:vrad}
\end{figure*}

The precision of the RVs ($\sim$ 0.3 km $\rm{s}^{-1}$) obtained from GES observations \citep{Jackson_2015} allows us to study the kinematic properties of the cluster samples. We use the RVs to determine the intrinsic RV dispersion of each cluster, $\sigma_c$, and the probability of a spectroscopic candidate to belong to the cluster. For each cluster we use the stars selected as spectroscopic candidates from the analysis in Sect. \ref{sec:member}. 

In Fig. \ref{fig:vrad} the RV distributions of each cluster are shown. We modeled these using a maximum likelihood technique developed by \cite{Cottaar_2012}\footnote{Available on-line at https://github.com/MichielCottaar/velbin.}. Briefly, this technique fits the observed distribution with a model that assumes that the intrinsic RV distribution is a Gaussian which is broadened by the orbital motions of unresolved binary systems and by uncertainties in the RV measurements. The broadening due to the binaries is modelled by assuming the same distribution of orbital parameters as found in solar-type field stars, namely, a log-normal distribution of the binary periods with a mean 5.03 and dispersion 2.28 in $\rm{log_{10}}$ days \citep{Raghavan_2010}, a power-law $ \frac{dN}{dq} \sim q^{0.25}$ for 0.1 < \textit{q} < 1 \citep{Reggiani_2013} for the secondary to primary mass ratio (\textit{q}), and a flat distribution of eccentricity between 0 and the maximum value $e_{max}$ defined in \cite{Parker_2009}.

Since the analysis in Sect. \ref{sec:member} excludes the obvious non-members, the sample of spectroscopic candidates will not be entirely clean of contaminating field stars. Therefore, we add a second broader Gaussian distribution to the model to account for their presence. In the model the properties of field populations are free parameters without boundaries. In the case of NGC 2547 we perform the fit with three distinct Gaussian populations to take into account the presence of the population B of young stars in the Vela OB2 associations found by \cite{Sacco_2015}.\\
Since the uncertainties on the RV measurements have been empirically calculated only for the GIRAFFE targets \citep{Jackson_2015}, we exclude the UVES targets from the fits. We perform three fits for each cluster, with the fraction of the binaries ($f_{bin}$) fixed at three different values: 0.2, 0.5, and 0.8. For all clusters, we model only the stars with RVs inside the range -90 $\leq$ RV $\leq$ 90 km $\rm{s^{-1}}$.
\begin{table}[!h]
	\caption{Best parameters from the fits of the RV distributions.}
	\label{tab:vrad}  
	\centering
	\tiny{
		\begin{tabular}{lccc}
			\midrule
			\midrule		
			Cluster & $f_{bin}$ &$\rm{v_c}$ & $\sigma_{c}$\\
			&  (\%) & (km $\rm{s^{-1}}$) & (km $\rm{s^{-1}}$) \\			
			\midrule
			IC 2602 & 0.2 & $17.65 \pm 0.18$ & $0.75 \pm 0.40$ \\
			IC 2602 & 0.5 & $17.63 \pm 0.16$ & $0.60 \pm 0.20$\\
			IC 2602 & 0.8 & $17.60 \pm 0.16$ & $0.45 \pm 0.20$\\
			\midrule
			IC 2391 & 0.2 & $15.04 \pm 0.19$ & $0.65 \pm 0.19$\\
			IC 2391 & 0.5 & $14.98 \pm 0.17$ & $0.53 \pm 0.17$\\
			IC 2391 & 0.8 & $15.02 \pm 0.18$ & $0.43 \pm 0.18$\\
			\midrule
			IC 4665 & 0.2 & $-13.83 \pm 0.16$& -\\
			IC 4665 & 0.5 & $-13.64 \pm 0.21$& < 0.5\\
			IC 4665 & 0.8 & $-13.69 \pm 0.21$& -\\
			\midrule
			NGC 2547 & 0.2 & $12.79 \pm 0.10$ & $0.79 \pm 0.11$\\
			NGC 2547 & 0.5 & $12.80 \pm 0.09$ & $0.63 \pm 0.09$\\
			NGC 2547 & 0.8 & $12.81 \pm 0.09$ & $0.51 \pm 0.11$\\
			\midrule
	\end{tabular}}
\end{table}

Table \ref{tab:vrad} shows the results of the fits. Both the central velocity $\rm{v_{c}}$ and the intrinsic dispersion $\sigma_{c}$ derived for a binary fraction of 0.2 and 0.8 are within the error bounds of the best values obtained for a fraction of 0.5 (within 1$\sigma$ for IC 2602 and IC 2391 and 2$\sigma$ for NGC 2547), therefore we will adopt the results obtained with a binary fraction set to 0.5 as the best values for the rest of the paper. Since the intrinsic RV dispersion for IC4665 is too small to be resolved with our data, we can only estimate an upper limit of $\sim 0.5~\rm km~s^{-1}$, which is slightly larger than the typical error of our RV measurements ($\sim$ 0.3 km $\rm{s}^{-1}$).

Our mean RV estimates are in agreement with the values found by previous works for IC 2602 and IC 2391 \citep{Marsden_2009}, for IC 4665 \citep{Jeffries_2009}, and for NGC 2547 \citep{Sacco_2015}.

Using the RVs of the spectroscopic candidates, we also estimate the probability that each of them is a true member. In particular, starting from the assumptions of our models, we can calculate the probability for a cluster member ($p_c(\rm{v_r})$) and a field star ($p_f(\rm{v_r})$) to have RV = $\rm{v_r}$ given the set of best fit parameters. Starting from these functions, the membership probability of a star is $p_{cl}(\rm{v_r})=p_c(\rm{v_r})$$/$$(p_c(\rm{v_r})$$+$$p_f(\rm{v_r}))$. 

\begin{table*}[!h]
	\centering
	\caption{Robustness of the fits assuming different conditions of the binary properties.}
	\label{tab:test_fit}  
	\tiny{
		\begin{tabular}{lccccc}
			\midrule
			\midrule		
			Cluster & $\sigma_{c}$ &  $\sigma_{\rm{log}P \, = \, 4.33 \, \textit{days}}$ & $\sigma_{\rm{log}P \, = \, 5.73 \, \textit{days}}$ & $\sigma_{f(e) \, \sim \, e^2}$ & $\sigma_{\frac{dN}{dq} \, = \, \rm{flat}}$\\
			& (km $\rm{s^{-1}}$) & (km $\rm{s^{-1}}$) & (km $\rm{s^{-1}}$) &  (km $\rm{s^{-1}}$) & (km $\rm{s^{-1}}$) \\			
			\midrule
			IC 2602 & $0.60 \pm 0.20$ & $0.61 \pm 0.22$ & $0.60 \pm 0.21$ & $0.62 \pm 0.21$ & $0.62 \pm 0.24$\\
			IC 2391 & $0.53 \pm 0.17$ & $0.51 \pm 0.16$ & $0.53 \pm 0.18$ & $0.49 \pm 0.16$ & $0.54 \pm 0.17$\\			
			NGC 2547 & $0.63 \pm 0.09$ & $0.62 \pm 0.09$ & $0.61 \pm 0.09$ & $0.64 \pm 0.09$ & $0.66 \pm 0.10$\\			
			\midrule
	\end{tabular}}
\end{table*}

\subsubsection{Assumptions on binary properties and robustness of fits}

Our model assumes that the properties of binaries are distributed as for solar mass stars in the the solar neighborhood. However, as discussed by \cite{Burgasser_2007,Raghavan_2010} and \cite{Duchene_2013}, binary properties probably change as a function of the stellar mass or may depend on the dynamical evolution of the star-forming region where they have been formed \citep[e.g.,][]{Marks_2011}. Therefore, we perform tests in order to investigate how the results depend on the assumed binary properties. Specifically, we calculate the best fit values assuming: a) a mean binary period a factor five lower and higher than that found for solar mass stars by \cite{Raghavan_2010}; b) a distribution of eccentricities in the form f(\textit{e}) $\sim e^2$ between 0 and $e_{max}$, instead of a flat distribution; c) a flat distribution for the mass ratio \textit{q} rather than the power-law defined by \cite{Reggiani_2013}. In the case of mass ratio \textit{q}, for the test we used the flat distribution since it is strongly supported by observational evidence \cite[e.g.,][for a review]{Mermilliod_1999, Patience_2002, Bender_2008, Duchene_2013}. Other distributions have been proposed in the literature, for example the random pairing distribution, where the smallest mass is randomly drawn from the mass distribution \cite[e.g.,][]{Kroupa_1995b}. However, the random pairing distribution has been ruled out both theoretically and observationally \citep{Kouwenhoven_2005,Kouwenhoven_2007a,Kouwenhoven_2007b,Kobulnicky_2007,Metchev_2008,Kouwenhoven_2009}. The results of our tests, reported in Table \ref{tab:test_fit}, show that our assumptions of the binary properties do not strongly affect our final results. Since we estimate an upper limit on the $\sigma_{c}$ of IC 4665, we do not consider this cluster in these tests. 

\subsection{Velocity dispersion from TGAS}
\label{ref:TGAS}
\begin{table}[!ht]
	\centering
	\caption{Velocity dispersion estimates obtained with the maximum likelihood procedure described in the text using the Nelder-Mead method, except for where indicated by the asterisk (where the Newton Conjugate Gradient method was employed). The quoted errors are obtained using the Cram\'er-Rao inequality.}
	\label{tab:tgas_analysis}
	\tiny{
		\begin{tabular}{ccccc}
			\midrule
			\midrule
			& IC 2602 & IC 2391 & IC 4665 & NGC 2547 \\ 
			\midrule
			\midrule
			$\mathrm{N_i}$ & 66  &  43 & 16 & 34\\
			$\sigma_v \, [\mathrm{km\,  s^{-1}}]$  & $0.48 \pm 0.04$ & 0.59 $\pm$ 0.06 & 0.18 $\pm$ 0.04 & 0.43 $\pm$ 0.08 \\
			$\mathrm{N_f}$ & 63 & 42 &  15 & 34\\
			$\sigma_v \, [\mathrm{km\,  s^{-1}}]$  & 0.20 $\pm$ 0.02 & 0.43 $\pm$ 0.05  & 0.03 $\pm$ 0.04 & 0.43 $\pm$ 0.08 \\
			$\sigma_{\perp} [\mathrm{km\,  s^{-1}}] $ & 0.32 $\pm$0.02 & 0.42 $\pm$ 0.05 &  0.10 $\pm$ 0.02& 0.60 $\pm $ 0.10 \\
			\midrule
			\midrule
			$\mathrm{N_{r < \rm{R_{GES}}, i}}$ &  38 & 22 & 10  & 17 \\
			$\sigma_v \, [\mathrm{km \, s^{-1}}]$ &  0.18 $\pm$ 0.02 & 0.20 $\pm$ 0.04 & 0.05 $\pm$ 0.03 & 0.24 $\pm$ 0.08 (*) \\
			$\mathrm{N_{r < \rm{R_{GES}}, f}}$ & 37 & 22 & 10 & 17	 \\
			$\sigma_v \, [\mathrm{km \, s^{-1}}]$ & 0.16 $\pm$ 0.02 & 0.20 $\pm$ 0.04 & 0.05 $\pm$ 0.03 &  0.24 $\pm$ 0.08 (*)\\
			$\sigma_{\perp}  [\mathrm{km\,  s^{-1}}]$ & 0.24 $\pm$ 0.02 & 0.30 $\pm$ 0.05 &  0.13 $\pm 0.03$ &  0.40 $\pm$ 0.10\\
			\midrule
			\midrule
			\multicolumn{5}{l}{
				\begin{minipage}{0.48\textwidth}
					\footnotesize{\textbf{Notes.} The first row lists the initial number of stars, from G17. The second row gives the values of $\sigma_v$ estimated using the stars reported in the first row. The third row gives the number of stars remaining after the exclusion procedure, and the fourth and fifth rows report $\sigma_v$ and $\sigma_{\perp}$. The second half of the table is the same as the first half, except for the initial number of stars. Row six indeed lists the number of stars within the radii from the cluster center reported in Table \ref{tab:total_mass}.}
			\end{minipage}}								
	\end{tabular}}
\end{table}

The clusters studied in this work have been investigated by \cite{Gaia_2017} (hereafter G17), who used the Tycho-\textit{Gaia} Astrometric Solution (TGAS) subset of the first \textit{Gaia} data release \cite[DR1,][]{Gaia_2016_2,Gaia_2016_1} to derive cluster memberships, mean parallaxes, and proper motion values. Parallaxes have also been determined by \cite{Randich_2017} who, for these clusters, found an excellent agreement.

There is not much overlap between the TGAS (exclusively brighter stars) and GES samples. On one side this can be considered as a limitation (for example, we have RVs for the GES stars, but we lack astrometry, and vice-versa), however it can also be seen as an opportunity to derive certain cluster properties in an independent way. We focus on the velocity dispersion of the four clusters, and on the comparison of the values obtained using the two samples.

To derive the velocity dispersion using the TGAS data, we apply the maximum likelihood procedure described in \cite{Lindegren_2000} (hereafter L00), in particular in Appendix A.4 of their paper, to the stars selected as members by G17\footnote{The python implementation of the procedure is available at: \url{https://github.com/eleonorazari/KinematicModelling}.}. Assuming that all the stars in a moving group share the same space velocity with a small isotropic internal velocity dispersion, L00 determine the group centroid space motion, the internal velocity dispersion, and the individual parallaxes for all members. The observables used by L00 are parallaxes and proper motions, which are modelled as random variables with a probability density function (PDF) depending on the model parameters. The model parameters are the cluster centroid space motion $v_0$ , the velocity dispersion $\sigma_v$ and the \textit{n} parallaxes of the \textit{n} stars, $\varpi$. They further assume that the observations are independent and unbiased.\\
The likelihood function is the product of the single PDFs of all the stars. The method requires that the model provides a statistically corrected description of the data. In particular, it must be applied to actual members of the cluster, or to the sources whose space motion agrees with the model. Outliers can be detected by computing a suitable goodness of fit statistic for each star in the solution. L00 named this quantity $g_i$ for each star with index \textit{i} (with \textit{i} = 1, ..., \textit{n}), and find that $g_i$ approximately follows a $\chi^2$ distribution. Therefore, for a given significance level, the star should be considered as a kinematic outlier if $g_i > g_{lim}$. For example, a 1 per cent significance level requires $g_{lim} \sim 14$. The outlier rejection procedure is iterative, where at each step the star with the largest $g_i$ is rejected from the sample. A new solution is then computed, including new $g_i$ values. The process is repeated until all $g_i < g_{lim}$.

Unfortunately, the internal velocity dispersion $\sigma_v$ is strongly underestimated by this method. The bias in the $\sigma_v$ estimate is probably related to the fact that an isotropic velocity dispersion is assumed for the cluster, while in practice only one component of this velocity can be measured astrometrically, i.e. the one perpendicular to the plane containing the line of sight and the centroid velocity vector (called $\eta_{\perp,i}$ by L00). L00 deal with the problem by using the proper motions residuals to compute the peculiar velocity components ($\eta_{\perp,i}$) and their observational uncertainties. Then, they compute an estimate of $\sigma_{\perp}$, and hence of $\sigma_v$, assuming isotropic dispersion. They test the method using Monte Carlo simulations, and conclude that $\sigma_{\perp}$ is in practice an unbiased estimate of $\sigma_v$.

We apply the likelihood maximization procedure described above first considering all the the stars identified as members by G17, then restricting the samples to the areas defined by the radii ($\rm{R_{GES}}$) shown in Table \ref{tab:mass_radii}. We derive the centroid space velocity for the four clusters, then we compute $\sigma_{\perp}$. The results are reported in Table \ref{tab:tgas_analysis}, together with their statistical errors\footnote{In practice, to maximize the likelihood we used the 'Nelder-Mead' and 'Newton-CG' methods, both supported by the \texttt{scipy.optimize.minimize} function.}.

\noindent
The estimated values for the velocity dispersion of the clusters analyzed in this study depend strongly on the number of stars considered. For example, the velocity dispersion of IC 2602 changes from $\sim 0.48$  km $\rm{s^{-1}}$ to $\sim 0.20$  km $\rm{s^{-1}}$ after the exclusion of three stars only. A similar trend can be observed also for the other clusters. Furthermore, changing the likelihood maximization method (see Table \ref{tab:tgas_analysis2} in appendix A) causes the velocity dispersion estimates to change slightly as well. For these reasons, the results reported in Table \ref{tab:tgas_analysis} and \ref{tab:tgas_analysis2} needs to be interpreted with care. In particular, the errors reported in Table \ref{tab:tgas_analysis} and \ref{tab:tgas_analysis2} correspond to the statistical errors, and do not take into account any systematic effect. A tentative estimate of the accuracy of the velocity dispersions obtained can be computed using half the difference between the velocity dispersion values obtained with the two different methods, considering the same number of stars (i.e. $\mathrm{N_{i}}$ and $\mathrm{N_{r < \rm{R_{GES}}, i}}$). In this way, we obtain systematic errors between $\sim 0.01$  km $\rm{s^{-1}}$ and $\sim 0.1$  km $\rm{s^{-1}}$, depending on the cluster. 

\subsection{Stellar mass and radii}
\label{ref:Masses}
\begin{figure*}[!t]
	\centering
	\includegraphics[width=0.99\linewidth]{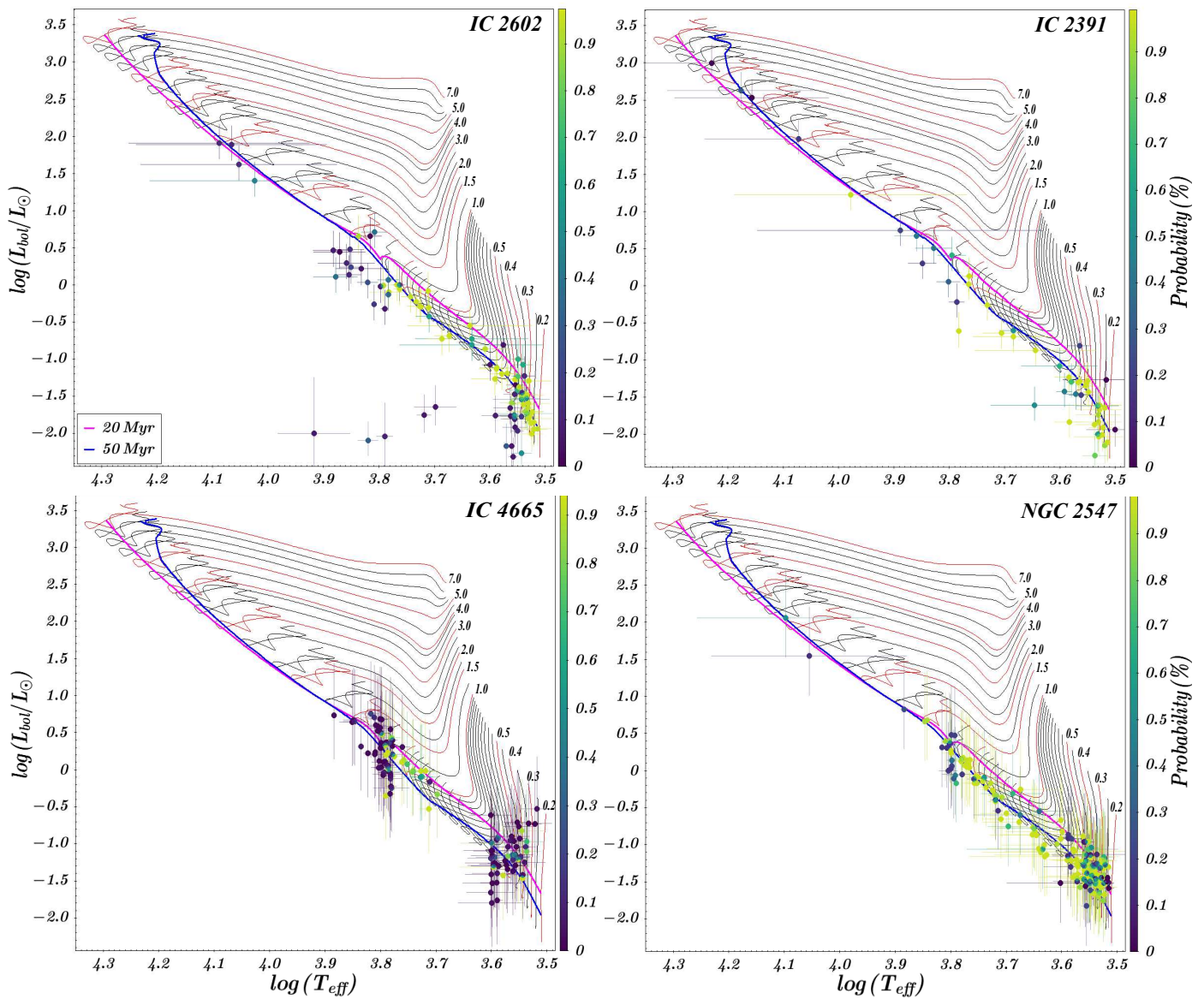}
	\caption{Hertzsprung-Russell diagram of sample of clusters. The colored filled dots represent GES targets retained after the procedure of cleaning from the obvious contaminants. The spectroscopic candidates have a different color depending on the probability of belonging to the cluster. The magenta and blue lines are the isochrones at 20 and 50 Myr, respectively, and the black and red lines are the PMS evolutionary tracks \citep{Tognelli_2011}.}
	\label{fig:HRD_selection_mem}
\end{figure*}

The analysis of the dynamics of clusters requires an estimation of its total mass and its half mass radius. As a first step, we calculated the mass of each spectroscopic candidate by interpolating the PMS evolutionary tracks developed by \cite{Tognelli_2011} at the positions of the stars in the HR diagram. We use effective temperatures measured from the GES spectra and luminosities estimated from the V magnitude or the J magnitude from the 2MASS catalogue, when the former is not available. To estimate luminosities from magnitudes, we correct for extinction, using the reddening values given in Table \ref{tab:clusters} and the extinction law from \cite{Savage_1979}; we apply bolometric corrections $BC_V$ and $BC_J$ derived by interpolating the relations reported in Table A5 of \cite{Kenyon_1995} at the stellar effective temperature; we convert relative bolometric magnitudes into luminosities adopting the distances reported in Table \ref{tab:clusters} and a solar bolometric magnitude $M_{\odot} = 4.74$. Figure \ref{fig:HRD_selection_mem} shows the HR diagram for each cluster, color-coded by membership probability. We note that the high probability spectroscopic candidates tend to be closer to the cluster sequence than low probability ones. This validates our approach to work.

Once we evaluate the mass of the spectroscopic candidates, we estimate the total mass of the cluster ($\rm{M}_{tot}$) using a general method. We take into account the stars within a magnitude range and we obtain the observed mass ($\rm{M}_{obs}$) adding up all star masses in the sample. Then, we need to divide $\rm{M}_{obs}$ by the level of completeness of the observations (Sect. \ref{sec:completeness}). This factor takes into account the fraction of potential cluster members within the magnitude range and within the area covered by the observations that have not been observed for technical reasons (e.g., the impossibility to allocate the fibers). Thereafter, we multiply $\rm{M}_{obs}$ for a factor 1.25, which takes into account the presence of binaries, under the assumption of a 50\% binary fraction and a mass ratio with a flat distribution. As last step, in order to estimate $\rm{M}_{tot}$, we need to multiply the observed mass by another factor that takes into account the fraction of the cluster mass in stars outside the magnitude range. This is calculated analytically, using the assumption that the mass function of the clusters follows a multi-power-law described in \cite{Kroupa_2001b} between 0.01 $\rm M_{\odot}$ and the mass of the most massive cluster star known in the literature. 

We consider two different samples of stars:
\begin{enumerate}[(a)]
	\item the stars that we selected as spectroscopic candidate with GES. Internally the GES sample, we also use two different approaches: (i) we consider only the spectroscopic candidates with a probability to be a member greater than 0.8, and (ii) we consider all spectroscopic candidates weighted by their corresponding $p_{cl}$;\\
	\item the stars identified as members by G17 within the area defined by $\rm{R_{GES}}$. For this sample, we assume a level of completeness of 100\% since TGAS is assumed to be complete.
\end{enumerate}	
Table \ref{tab:mass_radii} shows the magnitude range of GES, the $\rm{R_{GES}}$, and the derived completeness. 

\begin{table*}[!t]
	\centering
	\caption{Completeness, total mass and half mass radius of the four clusters calculated with the three different methods.}
	\label{tab:mass_radii}  
	\tiny{
		\begin{tabular}{cccccccc}
			\midrule
			\midrule		
			Cluster &  J Magnitude & $\rm{R_{GES}}$ & Completeness & $\rm{M_{tot,0.8}}$ & $\rm{M_{tot,w}}$ & $\rm{M_{tot,TGAS}}$ &  $r_{hm}$\\ 
			& range completeness & (pc) &  (\%) &($\rm{M}_{\odot}$) & ($\rm{M}_{\odot}$) & ($\rm{M}_{\odot}$) & (pc)\\ 				
			\midrule
			IC 2602 & 6.8 -- 12.0 & 4.13 & $\sim 25$ &$\sim 173$ & $\sim 244$ & $\sim 229$ & $\sim 1.56$ \\
			IC 2391 & 7.0 -- 12.8 & 2.55 & $\sim 25$ & $\sim 111$ & $\sim 151$ & $\sim 126$ & $\sim 0.98$ \\
			IC 4665 &  10.0 -- 16.0 & 4.47 & $\sim 65$ &$\sim 78$ & $\sim 96$ & $\sim 144$ & $\sim 1.19$\\
			NGC 2547 & 8.0 -- 15.5 & 3.18 & $\sim 75$ &$\sim 176$ & $\sim 201$ & $\sim 216$ & $\sim 0.80$ \\
			\midrule
			\midrule
			\multicolumn{8}{l}{
				\begin{minipage}{0.67\textwidth}
					\footnotesize{\textbf{Notes.} $\rm{M_{tot,0.8}}$, $\rm{M_{tot,w}}$, and $\rm{M_{tot,TGAS}}$ indicate the total masses calculated using the sample (a) with approach (i), the sample (a) with approach (ii), and the sample (b), respectively.}
			\end{minipage}}									
	\end{tabular}}
\end{table*}

The total masses found with the different samples are in agreement within a factor of $\sim 1.8$. In particular, the results show a good correspondence between the masses estimated through the spectroscopic candidates weighted by $p_{cl}$ ($\rm{M_{tot,w}}$) and those calculated with the TGAS sample ($\rm{M_{tot,TGAS}}$) within the GES region. This validates our results, since the masses are estimated starting from almost independent star samples (only a tiny fraction of them is in common). Therefore, we will adopt for the subsequent dynamical analysis $\rm{M_{tot,w}}$.

To study the dynamical properties of each cluster we need also an estimate of the half mass radius ($r_{hm}$). This radius is critically dependent on the presence of mass segregation in the clusters. Indeed, $r_{hm}$ gets smaller with increasing of level of mass segregation. Given that the GES magnitude range does not allow us the observation of very brighter (and massive) stars, it is difficult to take into account the presence of mass segregation with the GES data. Instead, in the TGAS sample there are the brighter stars of each cluster and the observations are spatially complete. Therefore, we use the stars identified as members by G17 within $\rm{R_{GES}}$. To correctly take into account the presence of mass segregation, we also consider the cluster members present in the literature outside the \textit{Gaia} magnitude range (stars with V $\lesssim 6$). We consider as $r_{hm}$ the radius that contains half of the mass given by the sum of masses of TGAS and literature stars. The $r_{hm}$ of the four clusters are listed in Table \ref{tab:mass_radii}.

\section{Discussion}

\subsection{GES versus TGAS velocity dispersion}

\begin{figure*}[!t]
	\centering
	\includegraphics[width=0.99\linewidth]{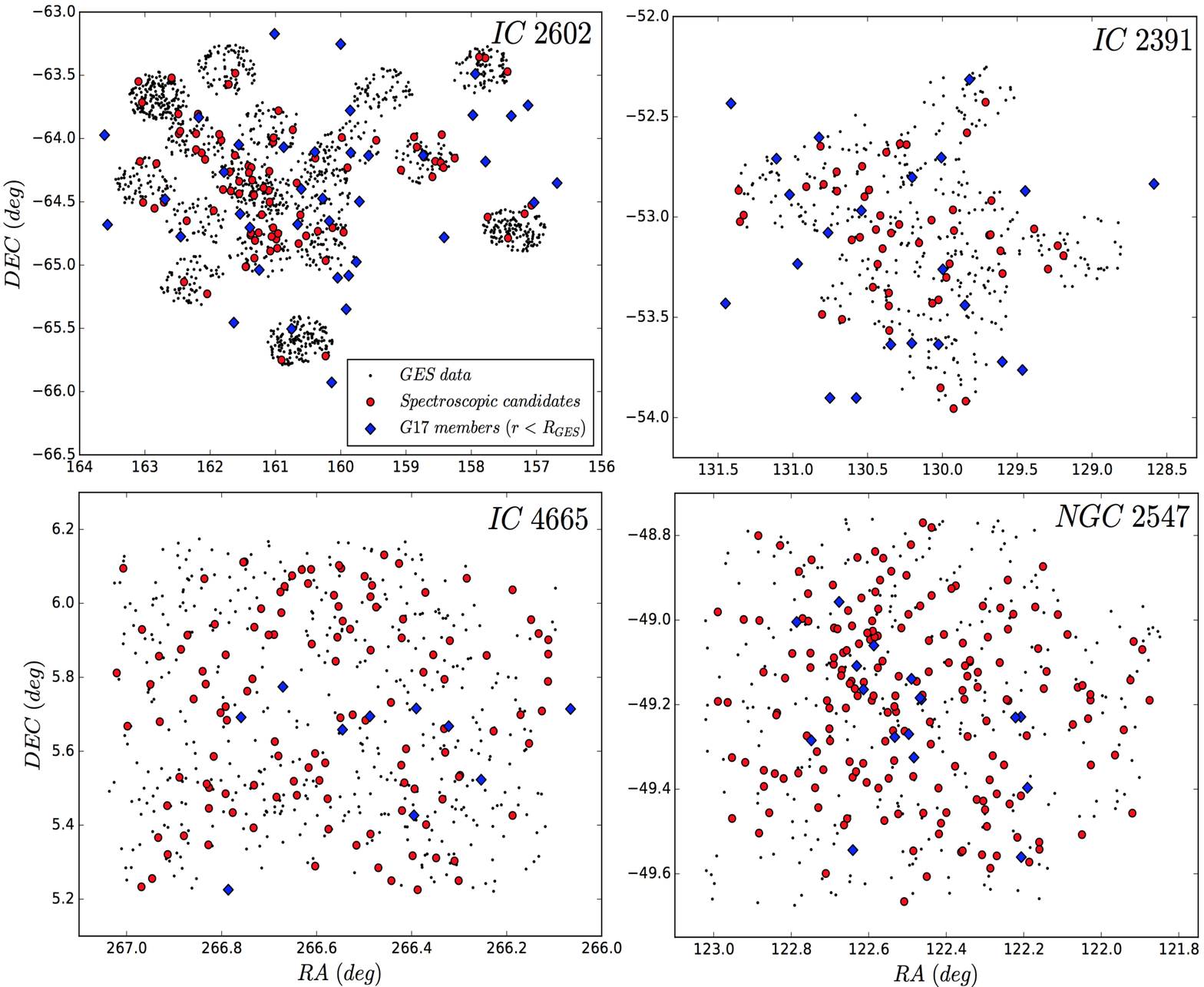}
	\caption{Spatial distribution of GES and TGAS stars in the four clusters. The black points are the stars observed with GES and the red filled dots are the spectroscopic candidates selected in Sect. \ref{sec:member}. The blue diamonds are the stars selected as members by G17 within the radius $\rm{R_{GES}}$. The stars identified with red dots and blue diamonds are those used to derive the velocity dispersions in GES and TGAS sample, respectively.}
	\label{fig:disposition}
\end{figure*}
In this section we compare the velocity dispersion obtained from the GES (see Sect. \ref{sec:RV}) and TGAS data (see Sect. \ref{ref:TGAS}). We decided to use the values of $\sigma_{\perp}$ estimated from the sample of G17 within the radius $\rm{R_{GES}}$ (last row of Table \ref{tab:tgas_analysis}), with an error given by the sum of statistical error and systematic error ($\sim 0.1$  km $\rm{s^{-1}}$). In Fig. \ref{fig:disposition}, we show, for the four clusters, the spatial distribution of the members selected by G17 from the TGAS catalogue and the spectroscopic candidates selected by GES. The G17 members of IC 2602 and IC 2391 within $\rm{R_{GES}}$ are uniformly distributed in the whole area of GES observations. Instead, for IC 4665 and NGC 2547, the TGAS members seem to cover only a section of the cluster area. This may be related to the larger distance of these clusters, indeed, the membership selection in G17 is based on position, parallaxes and proper motions, so it is strongly affected by distance. 

The velocity dispersion derived from the GES data are higher than those derived from the TGAS data but consistent within 2$\sigma$. Furthermore, we stress that the typical error uncertainty in proper motions in DR1 \citep{Gaia_2016_2} is $\sim 1$ mas $\rm{yr^{-1}}$, which corresponds to about 0.7 -- 1.7 km $\rm{s^{-1}}$ at the distance of the four clusters studied in this paper. Therefore, we are pushing the \textit{Gaia} data to their precision limit and their systematic errors need to be better investigated.\\
However, all four cluster analyzed in this work show the same trend, i.e. TGAS dispersions are smaller than GES ones.  The origin of this discrepancy could be due to two main explanations. The first is the presence of asymmetries of the system \citep{Baumgardt_2007}. In fact, to derive the velocity dispersions, in GES we use the radial velocity of stars while in TGAS we use the velocity perpendicular to the plane containing the line of sight. Anyway, it is unlikely that all clusters show the same trend of asymmetry. \\
The second is the energy equipartition. After the relaxation, a cluster tends to evolve towards the energy equipartition, where the more massive stars settle on the center of cluster and cede kinetic energy to the less massive ones. In this case, the velocity dispersion is related to the mass \textit{m} of stars as $\sigma(m) \propto m^{-0.5}$. So, we expect that the more massive stars are dynamically colder (i.e., smaller velocity dispersion). The four clusters in this work might already be relaxed (relaxation times 10 -- 30 Myr) and we found that the median mass of GES samples are smaller than that of TGAS samples by a factor between $\sim 2.5$ and $\sim 4$, depending on the cluster. Therefore, we expect differences in velocity dispersions between a factor $\sim 1.6$ and $\sim 2$, which is about what we found. Anyway, the presence of energy equipartition in star clusters is still very debated. \cite{Spera_2016} and \cite{Parker_2016b} noted that energy equipartition may not occur even after many two-body relaxation timescales.

These are still preliminary results and more data with better accuracy are needed. The second release of the \textit{Gaia} data, expected for April 2018, will include parallaxes and proper motions of low mass population of these clusters. We will therefore be able to investigate this discrepancy more thoroughly. In light of this, in the next section we will discuss only the results obtained with GES data.

\subsection{Effect of feedback on the cluster dissipation mechanism}

The main goal of this paper is to probe the dynamical state of four 20-50 Myr old clusters (IC 2602, IC 2391, IC 4665, and NGC 2547) in order to investigate the mechanism leading to cluster dispersion. In particular, determining if they are "supervirial" or "subvirial" is critically important. Indeed, according to the so called "residual gas expulsion" scenario 
\citep[e.g.][]{Kroupa_2001, Goodwin_2006}, young clusters become supervirial after feedback from massive stars sweeps out the gas that did not form stars. Otherwise, according to other models, the gas dispersion does not affect the virial ratio of the cluster and the dynamical interactions in the denser regions of a cluster drive the dynamical evolution \citep[e.g.][]{Kruijssen_2012,Parker_2016}.

We can understand if a cluster is supervirial by comparing the measured one-dimensional velocity dispersion $\sigma_c$ with the value derived analytically ($\sigma_{vir}$) from the cluster properties under the assumption of virial equilibrium, which is given by the equation: 

\begin{equation}
\label{eqn:virial}
\sigma_{vir}=\sqrt{\frac{\rm{M}_{tot} \, \, G}{\eta \, \, r_{hm}}}
\end{equation} 

\noindent where $r_{hm}$ is the half mass radius, $G$ is the gravitational constant, $\eta$ is a dimensionless factor, 
which depends on the cluster density profile and is approximately equal to 10 for a Plummer sphere profile 
\citep[e.g.][]{Spitzer_1987,Portegies-Zwart_2010} and $\rm{M}_{tot}$ is the cluster mass. In Table \ref{tab:total_mass} 
we report the velocity dispersion derived from the GES RVs (see Sect. \ref{sec:RV}) and from eqn. \ref{eqn:virial}. 

\begin{table}[!ht]
	\centering
	\caption{Properties of the four clusters.}
	\label{tab:total_mass}  
	\tiny{
		\begin{tabular}{cccccc}
			\midrule 
			\midrule		
			Cluster &$\rm{M}_{tot}$& $r_{hm}$& $\sigma_{c}$ & $\sigma_{vir}$ & $\rm{M_{dyn}}$\\ 
			&($\rm{M}_{\odot}$)& (pc)& (km $\rm{s^{-1}}$) &(km $\rm{s^{-1}}$) &($\rm{M}_{\odot}$)\\ 
			\midrule
			IC 2602 & $\sim 244$ & $\sim 1.56$ & $0.60 \pm 0.20$ &  $\sim 0.26$&1275\\
			IC 2391 & $\sim 151$& $\sim 0.98$&  $0.53 \pm 0.17$ &  $\sim 0.26$&485\\
			IC 4665 & $\sim 96$& $\sim 1.37$&  < 0.5 & $\sim 0.19$&-\\
			NGC 2547 & $\sim 201$& $\sim 0.80$& $0.63 \pm 0.09$ & $\sim 0.33$&720\\
			\midrule
	\end{tabular}}
\end{table}

The observed velocity dispersions are larger than the values calculated by assuming virial equilibrium by about a factor 
two, leading to the conclusion that three out of four clusters (except IC 4665 that has an upper limit on the velocity dispersion) are supervirial. We can rule out that this conclusion is due to errors on the estimates of the velocity dispersion $\sigma_{vir}$. In particular, uncertainties on the mass are lower than a factor 1.5 as shown in table 6; the half mass radius $r_{hm}$ could be underestimated in case of mass segregation, because it has been calculated using the more massive stars of the sample, but a larger $r_{hm}$ implies a smaller $\sigma_{vir}$, so it will support our conclusion on the virial ratio of the clusters. Finally, \cite{Elson_1987} and \cite{Fleck_2005} found that deviation of the density profile from a Plummer sphere can lead to a value of $\eta$ lower by a factor 2. However, considering that $\sigma_{vir}\propto \eta^{-1/2}$ this deviation cannot explain a discrepancy of a factor two. 

The presence of clusters in a supervirial state after gas expulsion has been predicted by several N-body simulations
\citep[e.g.,]{Bastian_2006, Baumgardt_2007} supporting the "residual gas expulsion scenario". In particular, \cite{Baumgardt_2007} suggest that -- after the gas that did not form stars is swept out -- clusters expand so the virial dispersion decreases and the virial ratio increases. Then, they return in a virial state only after the unbound stars are dispersed, which should occurs after about 20 and 40 crossing times. If we calculate the crossing time as $\sigma_{c}/r_{hm}$, the clusters studied in this paper have a dynamical age of about 20 - 30 crossing times, therefore, our results is in a good agreement with these simulations. However, we note that the crossing time used to track the cluster evolution in the simulations is calculated at cluster formation. We do not know the initial crossing time of these four clusters, but it is likely shorter than current one, so the evolution of these clusters could be slower than observed in the simulations. 

\cite{Parker_2016} performed N-body simulations of the cluster evolution assuming an initial spatial distribution that better resembles the hierarchical structure observed in young star forming region and investigated if the ratio $\sigma_{c}/\sigma_{vir}$ can be used to trace the dynamical state of a cluster. They found that clusters that are initially subvirial, or in global virial equilibrium but subvirial on local scale, relax to virial equilibrium after 25 - 50 crossing times. However, the measured ratio $\sigma_{c}$/$\sigma_{vir}$ would lead to the conclusions that they are supervirial. This apparent inconsistency originates by the fact that clusters are never fully relaxed but keep an imprint of early non-equilibrium even after several crossing times.

Finally, we point out that G17 found members up to 15 pc from the cluster center and outside the cluster radius considered in this paper. It is not clear if these distant stars are actual cluster members, unbound escaping stars or field stars with kinematic properties consistent with the cluster. Anyway, if we calculate total mass of the cluster and the half mass radius using the full G17 sample, we found similar virial velocity dispersions $\sigma_{vir}$, therefore, our conclusions would not change.

\section{Summary}

In this paper we analyzed the iDR4 internal products of the \textit{Gaia}-ESO survey to study the kinematical and dynamical properties of the young (age 20 -- 50 Myr) open clusters IC 2602, IC 2391, IC 4665, and NGC 2547. 

Using a gravity index, the lithium equivalent width, and the metallicity we derived a sample of candidate members for each cluster. Then, we used the RVs to derive the cluster intrinsic velocity dispersion, and membership probabilities for each candidate member. Photometry from the literature and the effective temperature from GES spectra have been used to estimate stellar masses 
and the total mass of each cluster, after correcting for the presence of binaries and completeness.\\
Furthermore, we independently derived the intrinsic velocity dispersion of the clusters from the astrometric parameters of cluster members in the TGAS catalogue.

On the basis of this analysis we obtained the following main results: 

\begin{itemize}
	\item[$\bullet$] The velocity dispersion measured from the RVs is higher than that measured from TGAS data. Given the masses of the stars in the GES and in the TGAS sample, this discrepancy would suggest that the system is relaxed and in a state of energy equipartition. However, given the limited numbers of cluster members in the TGAS sample and the error on astrometric parameters, we are not able to draw a firm conclusion. Important progresses will be possible very soon with the second \textit{Gaia} data release. \\

	\item[$\bullet$] The velocity dispersion measured with GES data is higher by about a factor two than what calculated by assuming virial equlibrium, given the masses of the clusters and the spatial distribution of their members. This result indicates that clusters are in supervirial state and two explanations are given to interpret it. The first is the "residual gas expulsion" scenario \cite[e.g.,][]{Kroupa_2001,Goodwin_2006}, which suggests that clusters became unbound after the "feedback" from massive stars swept out the gas which did not form stars. The second is that the observed velocity dispersion could be higher than the virial one because most stellar systems do not fully relax, even after 20 - 30 crossing times, as shown in N-body simulations of \cite{Parker_2016}.\\

	\item[$\bullet$] In each cluster we found many new high probability members and confirmed many of those known in the literature. New high probability members are extended across the whole area covered by GES observations, suggesting that these clusters could be more extended than previously thought.
	
\end{itemize}

When the observations of the \textit{Gaia}-ESO survey are completed and data from the second \textit{Gaia} data release will be available, we will able to study the kinematic of a larger sample of young clusters in a six dimensional space and solve the many open issues in this area of star formation.

\begin{acknowledgements}
The authors thank the referee, R. J. Parker, for his review. Based on data products from observations made with ESO Telescopes at the La Silla Paranal Observatory under programme ID 188.B-3002. These data products have been processed by the Cambridge Astronomy Survey Unit (CASU) at the Institute of Astronomy, University of Cambridge, and by the FLAMES/UVES reduction team at INAF/Osservatorio Astrofisico di Arcetri. These data have been obtained from the Gaia-ESO Survey Data Archive, prepared and hosted by the Wide Field Astronomy Unit, Institute for Astronomy, University of Edinburgh, which is funded by the UK Science and Technology Facilities Council.
This work was partly supported by the European Union FP7 programme through ERC grant number 320360 and by the Leverhulme Trust through grant RPG-2012-541. We acknowledge the support from INAF and Ministero dell' Istruzione, dell' Universit\`a' e della Ricerca (MIUR) in the form of the grant "Premiale VLT 2012" and by PRIN-INAF 2014. The results presented here benefit from discussions held during the Gaia-ESO workshops and conferences supported by the ESF (European Science Foundation) through the GREAT Research Network Programme. This work has made use of data from the European Space Agency (ESA) mission {\it Gaia} (\url{https://www.cosmos.esa.int/gaia}), processed by the {\it Gaia} Data Processing and Analysis Consortium (DPAC, \url{https://www.cosmos.esa.int/web/gaia/dpac/consortium}). Funding for the DPAC has been provided by national institutions, in particular the institutions participating in the {\it Gaia} Multilateral Agreement.
L.B. wishes to thank J. G. Fern\'{a}ndez-Trincado and B. Tang for the useful comments and G. Conte who worked at the production of Fig. 3. E.Z. wishes to thank A. Brown, T. Marchetti, and C.F. Manara for useful discussions. J.L.-S. acknowledges the Office of Naval Research Global (award no. N62909-15-1-2011) for partial support. F.J.E. acknowledges financial support from ASTERICS project (ID:653477, H2020-EU.1.4.1.1. - Developing new world-class research infrastructures).
\end{acknowledgements}

\bibliographystyle{aa}
\bibliography{bibliography}

\begin{appendix}
	\label{appendix}
	\section{Velocity dispersion using the Newton Conjugate Gradient maximization method}
	In this Section	we show the results of the radial velocity dispersions obtained with the TGAS data using the Newton Conjugate Gradient maximization method.
	\begin{table}[!h]
	\caption{Same as Table \ref{tab:tgas_analysis}, but using the Newton Conjugate Gradient maximization method.}
	\label{tab:tgas_analysis2}
	\vspace{0.1cm}
	\centering
	\tiny{
	\begin{tabular}{ccccc}
		\midrule
		\midrule
		& IC 2602 & IC 2391 & IC 4665 & NGC 2547 \\ 
		\midrule
		\midrule
		$\mathrm{N_i}$ & 66  &  43& 16 & 34\\
		$\sigma_c \, [\mathrm{km\,  s^{-1}}]$&$0.38 \pm 0.03$&0.34 $\pm$ 0.04&0.20 $\pm$ 0.05&0.40 $\pm$ 0.07 \\
		\midrule
		$\mathrm{N_f}$ & 59 & 40 &  15 & 34\\
		$\sigma_c \, [\mathrm{km\,  s^{-1}}]$&0.12 $\pm$ 0.02&0.10 $\pm$ 0.02&0.02 $\pm$ 0.06&0.40 $\pm$ 0.07 \\
		$\sigma_{ \perp} [\mathrm{km\,  s^{-1}}] $&0.25 $\pm$ 0.02& 0.18 $\pm$ 0.02&0.11 $\pm$ 0.02&0.60 $\pm$ 0.10 \\
		\midrule
		\midrule
		$\mathrm{N_{r <\rm{R_{GES}}, i}}$ &  38 & 22 & 10 & 17 \\
		$\sigma_c \, [\mathrm{km \, s^{-1}}]$&0.25 $\pm$ 0.03&0.28 $\pm$ 0.05 & 0.16 $\pm$ 0.05& 0.24 $\pm$ 0.08\\
		\midrule
		$\mathrm{N_{r <\rm{R_{GES}}, f}}$ & 37 & 22 & 10 & 17\\
		$\sigma_c \, [\mathrm{km \, s^{-1}}]$ & 0.15 $\pm$ 0.02&0.28 $\pm$ 0.04 & 0.16 $\pm$ 0.03&0.24 $\pm$ 0.08 \\
		$\sigma_{\perp}  [\mathrm{km\,  s^{-1}}]$&0.25 $\pm$ 0.02& 0.30 $\pm$ 0.05&0.12 $\pm$ 0.02&0.37 $\pm$ 0.09\\
		\midrule
		\midrule
	\end{tabular}}
	\end{table}
\end{appendix}

\end{document}